\documentclass[acmsmall]{acmart}
\AtBeginDocument{%
  \providecommand\BibTeX{{%
    \normalfont B\kern-0.5em{\scshape i\kern-0.25em b}\kern-0.8em\TeX}}}

\setcopyright{acmcopyright}
\copyrightyear{2024}
\acmYear{2024}
\acmDOI{10.1145/3637841}
\acmJournal{CSUR}
\acmVolume{56}
\acmNumber{5}
\acmArticle{130}
\acmMonth{1}

\usepackage{tikz}
\usepackage{array}
\usepackage{multirow}
\usepackage{makecell}
\usepackage{enumitem}
\usepackage{hyperref}
\usepackage{amsmath,amsfonts,bm}
\usepackage[ruled,vlined,noend]{algorithm2e}
\usepackage[inkscapelatex=false]{svg}
\usepackage{marvosym}
\DeclareMathOperator*{\onehot}{one-hot}
\DeclareMathOperator*{\id}{\texttt{id}}
\DeclareMathOperator*{\argmaxA}{arg\,max}
\DeclareMathOperator*{\R}{\mathbb{R}}

\begin{document}
\title{Embedding Compression in Recommender Systems: A Survey}

\author{Shiwei Li}
\authornote{Shiwei Li and Huifeng Guo contributed equally to this research.}
\email{lishiwei@hust.edu.cn}
\orcid{0000-0002-7067-0275}
\affiliation{
  \institution{Huazhong University of Science and Technology}
  \city{Wuhan}
  \country{China}
  \postcode{430074}
}

\author{Huifeng Guo}
\authornotemark[1]
\email{huifeng.guo@huawei.com}
\orcid{0000-0002-7393-8994}
\affiliation{
  \institution{Huawei Noah's Ark Lab}
  \city{Shenzhen}
  \country{China}
  \postcode{518129}
}

\author{Xing Tang}
\authornote{This work was done when Xing Tang worked at Huawei Noah's Ark Lab.}
\email{xing.tang@hotmail.com}
\orcid{0000-0003-4360-0754}
\affiliation{
  \institution{Tencent}
  \city{Shenzhen}
  \country{China}
  \postcode{518054}
}

\author{Ruiming Tang}
\email{tangruiming@huawei.com}
\orcid{0000-0002-9224-2431}
\author{Lu Hou}
\email{houlu3@huawei.com}
\orcid{0000-0002-4694-1821}
\affiliation{
  \institution{Huawei Noah's Ark Lab}
  \city{Shenzhen}
  \country{China}
  \postcode{518129}
}

\author{Ruixuan Li}
\email{rxli@hust.edu.cn}
\authornote{Ruixuan Li and Rui Zhang are the corresponding authors.}
\orcid{0000-0002-7791-5511}
\affiliation{
  \institution{Huazhong University of Science and Technology}
  \city{Wuhan}
  \country{China}
  \postcode{430074}}
  
\author{Rui Zhang}
\email{rayteam@yeah.net}
\authornotemark[3]
\orcid{0000-0002-8132-6250}
\affiliation{
  \institution{ruizhang.info}
  \country{China}}
\renewcommand{\shortauthors}{Shiwei Li and Huifeng Guo, et al.}


\begin{abstract}
To alleviate the problem of information explosion, recommender systems are widely deployed to provide personalized information filtering services. Usually, embedding tables are employed in recommender systems to transform high-dimensional sparse one-hot vectors into dense real-valued embeddings. However, the embedding tables are huge and account for most of the parameters in industrial-scale recommender systems. In order to reduce memory costs and improve efficiency, various approaches are proposed to compress the embedding tables. In this survey, we provide a comprehensive review of embedding compression approaches in recommender systems. We first introduce deep learning recommendation models and the basic concept of embedding compression in recommender systems. Subsequently, we systematically organize existing approaches into three categories, namely low-precision, mixed-dimension, and weight-sharing, respectively. Lastly, we summarize the survey with some general suggestions and provide future prospects for this field. 
\end{abstract}
\maketitle
\section{Introduction}
To alleviate the problem of information explosion, recommender systems~\cite{mm_23,rs_survey_2023} are extensively deployed to provide personalized information filtering services, including online shopping~\cite{din}, advertising systems~\cite{ftrl} and so on. Meanwhile, deep learning techniques have shown impressive capabilities in capturing user preferences for candidate items.
Thereupon, both the industry and research communities have proposed a variety of deep learning recommendation models (DLRMs) to enhance the performance of recommender systems, such as Wide \& Deep~\cite{widedeep} in Google Play, DIN~\cite{din} in Alibaba and DeepFM~\cite{DeepFM} in Huawei. 

\subsection{Deep Learning Recommendation Models} \label{sec:dlrms}
Recommender systems are utilized for a diverse range of tasks, such as candidate item matching~\cite{item_match}, click-through rate (CTR) prediction~\cite{feat_inter_aaai,feat_inter_kdd}, and conversion rate (CVR) prediction~\cite{cvr}. 
For each of these tasks, the employed DLRMs have undergone meticulous design processes to ensure optimal performance. 
However, without loss of generality, most DLRMs follow the \textit{embedding table} and \textit{neural network} paradigm~\cite{dcn,dcn_v2,DeepFM,simplex}, despite the specific design of the \textit{neural network} component may vary across different model architectures. 

\begin{figure}[h]
\centering
\includegraphics[scale=0.45]{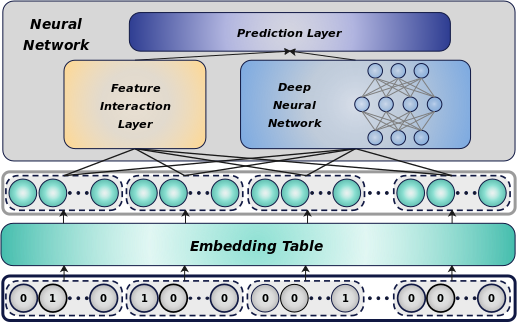}
\caption{The \textit{embedding table} and \textit{neural network} paradigm of deep learning recommendation models (DLRMs). Note that the \textit{neural network} component may vary in different model architectures, here we only present the \textit{neural network} with the classic dual tower architecture as an example.}\label{fig:embnn}
\end{figure}

As illustrated in Figure~\ref{fig:embnn}, the {embedding table} is responsible for converting input rows into dense embedding vectors. 
It is worth noting that the input rows of DLRMs typically consist of categorical features, which are encoded as high-dimensional one-hot vectors. 
Each category feature will be referred to as feature for short, and all features under the same category form a feature field. 
Generally, each feature is associated with a unique embedding stored in the embedding table $\mathbf{E}\in \mathbb{R}^{n\times d}$, where $n$ denotes the  total number of features and $d$ denotes the embedding dimension. 

On the other hand, the {neural network} is primarily engaged in interacting, processing, and analyzing feature embeddings, along with making predictions. 
Recent studies~\cite{DeepFM,ccpm,fgcnn,afm,autofeature,sif} have consistently focused on optimizing the feature extraction capabilities of the neural networks. 
For example, \cite{DeepFM,dcn_v2} utilize product operators to model the feature interactions between different feature fields. \cite{ccpm,fgcnn} employ convolutions on embeddings to capture feature interactions of arbitrary order. \cite{afm} introduces an additional attention network to assign varying importance to different feature interactions. Additionally, \cite{autofeature,sif} automatically search for suitable interaction functions using AutoML techniques~\cite{automl_survey}. In this manuscript, we do not delve into the detailed design of the {neural network} component. Instead, we recommend referring to the works~\cite{CTR_survey,rs_survey_2019,rs_survey_2023} for a comprehensive understanding of the neural networks used in DLRMs.

Despite incorporating various intricate designs, the neural network usually entails relatively shallow layers and a limited number of model parameters. 
In contrast, the embedding table occupies the vast majority of model parameters. 
Especially in industrial-scale recommender systems, where there are billions or even trillions of categorical features, the embedding table may take hundreds of GB or even TB to hold~\cite{SFCTR}. 
For example, the size of embedding tables in Baidu's advertising systems reaches 10 TB~\cite{Int16}. 
As the scale of recommender systems perpetually expands, the continuous growth in the number of features will bring greater storage overhead.

\subsection{Embedding Compression in Recommender Systems} \label{sec:emb_comp}

In addition to increasing storage overhead, larger embedding tables will also result in higher latency during table lookup~\footnote[1]{The process of retrieving an embedding from the embedding table based on the input feature or index.}, which will reduce the efficiency of model training and inference. Therefore, to deploy the DLRMs with large embedding tables in real production environment efficiently and economically, it is necessary to compress their embedding tables. 

However, embedding compression in DLRMs differs significantly from model compression in other fields, such as Computer Vision (CV)~\cite{DL_Compress} and Natural Language Processing (NLP)~\cite{NLP_compress}. 
These differences primarily manifest in three aspects: model architectures, properties of input data, and model size. 
Firstly, vision models and language models are usually very deep neural networks stacked by fully-connected layers, convolutional layers, or transformers. 
Consequently, compression methods designed for these models focus on compressing the aforementioned modules rather than embedding tables.
In contrast, DLRMs are typically shallow models, with the majority of parameters concentrated in the embedding tables. 
Secondly, the input data of vision models and language models are usually images and texts, inherently containing abundant visual and semantic information that can be leveraged for model compression.
For example, \cite{Basis_vector_word} use the semantic information as a prior knowledge to compress the word embeddings, while \cite{ghostnet} exploit the similarity between feature maps derived from image inputs to compress convolution kernels. 
However, in recommender systems, there is generally limited visual or semantic information available. 
Fortunately, DLRMs possess unique properties in the input data that can facilitate embedding compression. 
Specifically, categorical features are organized in feature fields and often follow a highly skewed long-tail distribution, with varying numbers of features in different fields.
We can compress embedding tables based on feature frequency and field size. 
Thirdly, embedding tables of DLRMs are usually hundreds or even thousands of times larger than vision models or language models \cite{CpRec}, which presents a more challenging and necessary task for compression.

\begin{figure}[t]
\begin{center}
\begin{tikzpicture}
[auto, block/.style ={rectangle, draw=black, thick, align=left, rounded corners, minimum width=5em, minimum height=1.5em, text width=18em}, line/.style ={draw, thick, -latex',shorten >=2pt}]
\tikzstyle{every node}=[font=\fontsize{6}{8}\selectfont]
\draw 
(0, -3.5)   node[block, align=center, text width=5em] (1) {\textbf{Embedding}\\ \textbf{Compression}}
(3, -1.0)   node[block, align=center, text width=5em] (11){\textbf{Low-precision}}
(3, -3.5)   node[block, align=center, text width=5em] (12){\textbf{Mixed-dimension}} 
(3, -6.0)   node[block, align=center, text width=5em] (13){\textbf{Weight-sharing}};

\draw
(8, -0.4)   node[block] (111){\begin{tabular}{l}\textbf{Binarization:}\\
                    \cite{Two-stage1}, \cite{Two-stage2}, DCF~\cite{DCF}, DCMF~\cite{DCMF}, DPR~\cite{DPR}, DFM~\cite{DFM} \\
                    CIGAR~\cite{CIGAR}, HashGNN~\cite{HashGNN}, L$^2$Q-GCN~\cite{L2q} \end{tabular}}
(8, -1.45)   node[block] (112){\begin{tabular}{l}\textbf{Quantization:}\\
                    \cite{Post}, \cite{Mixed-Precision}, \cite{Int16}, ALPT~\cite{alpt} \end{tabular}};
\draw
(8, -2.45)   node[block] (121){\begin{tabular}{l}\textbf{Rule-based Approaches:} \\  
                    MDE~\cite{MDE}, CpRec~\cite{CpRec} \end{tabular}}
(8, -3.5)   node[block] (122){\begin{tabular}{l}\textbf{NAS-based Approaches:}\\
                    NIS~\cite{NIS}, ESAPN~\cite{ESAPN}, AutoIAS~\cite{AutoIAS}, \\
                    AutoEmb~\cite{AutoEmb}, AutoDim~\cite{AutoDim},  RULE~\cite{Rule}, OptEmbed~\cite{OptEmbed} \end{tabular}}
(8, -4.55)   node[block] (123){\begin{tabular}{l}\textbf{Pruning:}\\ 
                    DeepLight~\cite{DeepLight}, DNIS~\cite{DNIS}, AMTL~\cite{AMTL}, PEP~\cite{PEP}, SSEDS~\cite{SSEDS} \end{tabular}};
\draw
(8, -5.55)   node[block] (131){\begin{tabular}{l}\textbf{Hashing:}\\ 
                    QR~\cite{QR}, MEmCom~\cite{MEmCom}, BCH~\cite{Binary}, FDH~\cite{Double}, LMA~\cite{LMA}, ROBE~\cite{ROBE} \end{tabular}} 
(8, -6.45)   node[block] (132){\begin{tabular}{l}\textbf{Vector Quantization:}\\ 
                    Saec~\cite{Saec}, MGQE~\cite{MGQE}, xLightFM~\cite{xLightFM}, LightRec~\cite{LightRec}, LISA~\cite{Linear} \end{tabular}}
(8, -7.35)   node[block] (133){\begin{tabular}{l}\textbf{Decomposition:}\\ 
                    MLET~\cite{Multi}, ANT~\cite{Ant}, DHE~\cite{DHE}, TT-Rec\cite{TT}, LLRec~\cite{TT-KD}, \cite{STT} \end{tabular}};

\draw (1.east) -- (11.west);
\draw (1.east) -- (12.west);
\draw (1.east) -- (13.west);
\draw (11.east) -- (111.west);
\draw (11.east) -- (112.west);
\draw (12.east) -- (121.west);
\draw (12.east) -- (122.west);
\draw (12.east) -- (123.west);
\draw (13.east) -- (131.west);
\draw (13.east) -- (132.west);
\draw (13.east) -- (133.west);
\end{tikzpicture}
\end{center}
\caption{Summary of representative studies on embedding compression in recommender systems.} \label{fig:summary}
\end{figure}
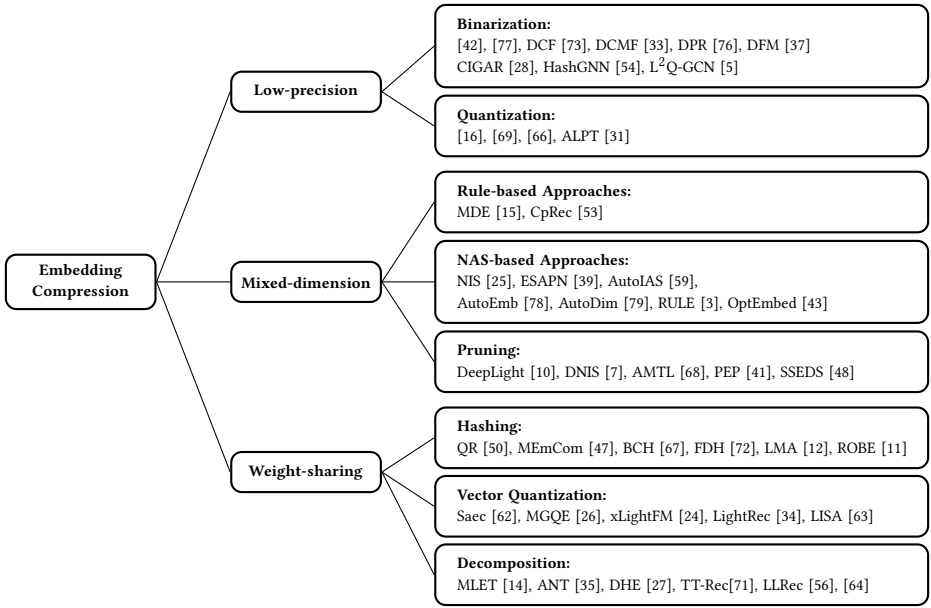

Recently, embedding compression has gained increasing attention in recommender systems, leading to the development and application of various embedding compression techniques for DLRMs. 
However, there is currently no comprehensive survey summarizing the methods employed for embedding compression. Therefore, the primary objective of this paper is to review and summarize representative research in this field. 
The embedding table can be regarded as a matrix with three dimensions, that are the precision of weights, the dimension of embeddings, and the number of embeddings. 
To this end, we summary the embedding compression methods into three categories according to the dimensions they compress, as illustrated in Figure~\ref{fig:summary}. 
Firstly, \textbf{low-precision} methods reduce the memory of each weight by decreasing its bit width. According to the size of bit width and its corresponding advantages, we further divide the low-precision methods into binarization and quantization. 
Secondly, \textbf{mixed-dimension} methods reduce the memory of specific embeddings by decreasing their dimensions and using mixed-dimension embeddings. According to the techniques of determining the embedding dimension for different features, we categorize the mixed-dimension methods into rule-based approaches, NAS-based approaches, and pruning. 
Thirdly, \textbf{weight-sharing} methods reduce the actual parameters of the embedding table by sharing weights among different embeddings. 
Considering that the number of features is given by the dataset, a solution to reduce the number of embeddings is to reuse embeddings among features. Furthermore, we generalize the sharing to the weight level and define the weight-sharing methods as generating embeddings with shared weights. 
According to the way embeddings are generated, we categorize the mixed-dimension methods into hashing, vector quantization, and decomposition. 
We will introduce the three primary categories in Sections~\ref{sec:lp},~\ref{sec:md} and ~\ref{sec:ws}, respectively. 

Note that embeddings are fed into the neural network as representations of categorical features and form the foundations of DLRMs. Therefore, when compressing embeddings, it may affect the model performance on many aspects, including model accuracy, inference efficiency, training efficiency, and training memory usage. We will discuss the pros and cons of different methods regarding these metrics at the end of each section. In Section~\ref{sec:summary}, the survey is summarized, providing general suggestions for different scenarios and discussing future prospects for this field.

\section{Low-Precision}\label{sec:lp} 
As we all know, embedding weights are typically stored in the format of FP32~\footnote[2]{The short of single-precision floating-point format.} which occupies 32 bits.
To reduce the storage of each weight, low-precision approaches are developed to represent a weight with fewer bits. 
In particular, according to the bit width of weights, low-precision approaches can be further divided into \textbf{binarization} and \textbf{quantization}.

\subsection{Binarization}
Binarization is to compress a full-precision weight into a binary code that only occupy 1 bit. 
It is widely used in the embedding-based similarity search of recommender systems~\cite{CIGAR, HashGNN}, 
since the binary embeddings have two distinct advantages compared to the full-precision ones: 
(1) less memory or disk cost for storing embeddings; 
(2) higher inference efficiency as the similarity (i.e., inner product) between binary embeddings can be calculated more efficiently through the Hamming distance, 
which has been proved in~\cite{DCF}. 

\cite{Two-stage1,Two-stage2} pioneered to obtain  binary embeddings in a two-stage (i.e, post-training) manner. 
Specifically, they first learn a full-precision embedding table while ignoring the binary constraints, 
and then perform binarization (e.g., $sign(x)$) on the full-precision embeddings to get binary embeddings. 
However, the binarization procedure is not in the training process and thus cannot be optimized by minimizing the training objective, 
which will bring large irreparable errors and fail to meet an acceptable accuracy. 
To reduce accuracy degradation, subsequent works have focused on end-to-end approaches to learn the binary embeddings during training.

\begin{figure}[h]
\centering
\includegraphics[scale=0.54]{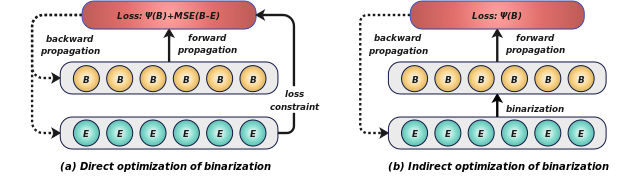}
\caption{End-to-end optimization paradigms of learning binary embeddings. $B$ (yellow) and $E$ (green) represent the binary and full-precision weights, respectively. 
$Loss$ (red) is the training loss where $\Psi(\cdot)$ is the objective function and $\text{MSE}(\cdot)$ is the mean-square-error function. }\label{fig:binarization}
\end{figure}

As shown in Figure~\ref{fig:binarization}, recent works typically learn binary embeddings following two optimization paradigms, {namely {direct optimization and indirect optimization}}. 
As Figure~\ref{fig:binarization}{(a)} shows, in the direct optimization of binarization, 
the binary embeddings are maintained as part of the model parameters and will be optimized directly by the training loss. 
For example, to improve the efficiency of  Collaborative Filtering (CF), DCF~\cite{DCF} learns a binary embedding table $\mathbf{B}\in {\{\pm 1\}}^{n\times d}$.  
To maximize the information encoded in each binary embedding, DCF further adds a balance-uncorrelation constraint to $\mathbf{B}$ 
(i.e., $\mathbf{B}^T\mathbf{1}=\mathbf{0},\mathbf{B}^T \mathbf{B}=n\mathbf{I}$), where $\mathbf{I}$ is an identity matrix. 
However, it is NP-hard to optimize the binary embeddings with such constraint. 
To resolve this problem, DCF also maintains a full-precision embedding table $\mathbf{E}\in \R^{n\times d}$ with the same balance-uncorrelation constraint. 
The constraint of $\mathbf{B}$ is then replaced by adding the mean-square-error (MSE) of $(\mathbf{B}-\mathbf{E})$ to the objective function. 
During training, DCF will update $\mathbf{B}$ and $\mathbf{E}$ alternatively through different optimization algorithms.
Specifically, $\mathbf{B}$ is updated by Discrete Coordinate Descent (DCD) and $\mathbf{E}$ is updated with the aid of Singular Value Decomposition (SVD). 
This optimization paradigm has been widely used to learn binary embeddings in recommender systems, such as DPR~\cite{DPR}, DCMF~\cite{DCMF}, DFM~\cite{DFM}. 
DPR changes the objective function of DCF (i.e., rating prediction) to personalized items ranking.
DCMF and DFM extends this binarization paradigm to Content-aware Collaborative Filtering~\cite{Content} and Factorization Machine (FM)~\cite{FM}, respectively.

As shown in Figure~\ref{fig:binarization}{(b)}, another paradigm is the {indirect optimization}, where the binary embeddings $\mathbf{B}$ are generated from full-precision embeddings $\mathbf{E}$ on the fly and will be optimized indirectly by optimizing $\mathbf{E}$.  
However, it is infeasible to optimize $\mathbf{E}$ by the standard gradient descent as the gradients of the binary operations (e.g., $sign(x)$) are constantly zero.
To solve this problem, CIGAR~\cite{CIGAR} replaces $sign(x)$ with the scaled tanh function $tanh(\alpha x)$ 
as $\lim_{\alpha  \to \infty}tanh(\alpha x) = sign(x)$ and $tanh(\alpha x)$ has better differential property. 
In the early stages of training, a smaller value of $\alpha$ is utilized to yield superior representations, and as the training progresses, its value gradually increases to approximate $sign()$. 
Another way to solve the non-propagable gradient is straight-through-estimator (STE)~\cite{STE}, which treats some operations as identity maps during backpropagation. HashGNN~\cite{HashGNN} employs the STE variant of $sign()$ and thus updating $\mathbf{E}$ with the gradients of $\mathbf{B}$. 
However, the huge gap between $\mathbf{B}$ and $\mathbf{E}$ will cause an imprecise update for $\mathbf{E}$. 
To solve this issue, HashGNN further develops a dropout-based binarization. 
Specifically, $\hat{{\mathbf{E}}}=(1-\mathbf{P})\odot \mathbf{E}+\mathbf{P}\odot \mathbf{B}$ will be fed into the following networks, 
where $\mathbf{P}\in\{0,1\}^{n\times d}$ and $\odot$ is the element-wise product. 
Each element in $\mathbf{P}$ is a Bernoulli random value with probability $p$. 
During backpropagation, only the embeddings that are binarized will be updated through STE and the rest will be updated by standard gradient descent. 
To ensure convergence, HashGNN adopts a small value for $p$ in the initial training phase, gradually increasing it as the training progresses. 
Similarly, \text{L$^2$Q-GCN}~\cite{L2q} uses STE to optimize the full-precision embeddings, while introducing a positive scaling factor $s=mean(|\bm{e}|)$ for each binary embedding to enhance its presentation capability, where $\bm{e}$ is the full-precision embedding. The comparison of the above three methods is summarized in Algorithm~\ref{alg:binary}.
\begin{algorithm}[htbp]
  \SetAlgoLined
  \DontPrintSemicolon
  \KwIn{a full-precision embedding $\bm{e}$.} 
  \KwOut{the output embedding $\hat{\bm{e}}$.\tcp*[r]{$\hat{\bm{e}}$ will be fed into following networks.}}
  \SetKwFunction{FC}{CIGAR}
  \SetKwFunction{FH}{HashGNN}
  \SetKwFunction{FL}{\text{L$^2$Q-GCN}}
  \SetKwProg{Fn}{Func}{:}{}
  
  \Fn{\FC{$\bm{e}$}}{
    $\hat{\bm{e}}=tanh(\alpha \cdot \bm{e})$ \tcp*[r]{$\alpha$ will increase as training progresses.}
  }

  \Fn{\FH{$\bm{e}$}}{
    $\bm{b}=sign\_ste(\bm{e})$  \tcp*[r]{$sign\_ste()$ is the STE variant of $sign()$.}

    $\bm{p} := \{0, 1\}^d$ \tcp*[r]{sample from Bernoulli distribution with probability $p$.} 
    
    $\hat{\bm{e}}=(1-\bm{p})\odot \bm{e}+\bm{p}\odot \bm{b}$\tcp*[r]{$p$ will increase as training progresses.}
  }

  \Fn{\FL{$\bm{e}$}}{
    $\bm{b}=sign\_ste(\bm{e})$ 
    
    $\hat{\bm{e}}=mean(|\bm{e}|) \cdot\bm{b}$ 
  }
  \caption{Comparison between CIGAR~\cite{CIGAR}, HashGNN~\cite{HashGNN} and \text{L$^2$Q-GCN}~\cite{L2q}.~\label{alg:binary}}
\end{algorithm}

\subsection{Quantization}
Although binarization has better efficiency and less memory cost at the inference stage, 
it may lead to a significant drop of accuracy, which is not acceptable in several scenarios.
As Cheng et al.~\cite{widedeep} claim, even $0.1\%$ decrease of the prediction accuracy may result in large decline in revenue. 
To trade off the memory cost and the prediction accuracy, quantization is used to represent each weight with a multi-bit integer. 

Quantization is the mapping of a 32-bit full-precision weight to an element in the set of quantized values $\mathbb{S}=\{ q_0, q_1, ...,q_k \}$, 
where $k=2^s-1$ and $s$ is the bit width. 
The most commonly used quantization function is uniform quantization, where the quantized values are uniformly distributed. 
Specifically, the step size $\Delta = q_{i}-q_{i-1}$ remains the same for any $i \in [1,k]$.
Let $w$ be a value clipped into the range $[q_0, q_k]$, we can quantize it into an integer as $\hat{w}=rd((w-q_0)/ \bigtriangleup)$, 
where $rd(x)$ rounds $x$ to an adjacent integer. 
The integer $\hat{w}$ will be de-quantized into a floating-point value $(\hat{w}\times\bigtriangleup+q_0)$ when used.  
Existing work on embedding quantization either performs post-training quantization or trains a quantized embedding table from scratch. 
\begin{figure}[h]
\centering
\includegraphics[scale=0.54]{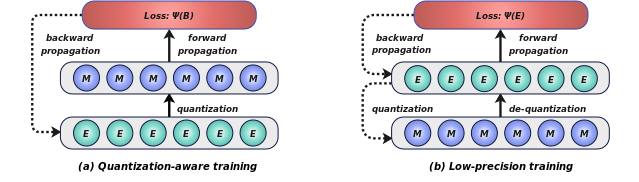}
\caption{Training frameworks of quantization. $Loss$ (red) is the training loss.  $M$ (purple) and $E$ (green) represent the integer and full-precision weights, respectively. 
}\label{fig:quantization}
\end{figure}

Guan et al.~\cite{Post} studies post-training quantization (PTQ) on the embedding tables and proposes a uniform and a non-uniform quantization algorithm. 
Specifically, in the uniform quantization, they maintain a quantization range for each embedding and find the best quantization range by a greedy search algorithm. 
In the non-uniform quantization, they divide similar embeddings into groups and apply k-means clustering on the weights to produce a codebook for each group.
The weights in each group will be mapped to the index of a value in the corresponding codebook. 
These two algorithms improve the accuracy of PTQ, however, they still suffer from accuracy degradation.

To further reduce accuracy degradation, recent works~\cite{Int16, Mixed-Precision} learn quantized weights from scratch. 
Unlike the well-known quantization-aware training (QAT)~\cite{lsq,lsq+}, 
\cite{Int16, Mixed-Precision} use another quantization training framework to exploit the sparsity of the input data, which we term low-precision training (LPT). 
As Figure~\ref{fig:quantization}(a) shows, QAT quantizes the full-precision weights in the forward pass and updates the full-precision weights with the gradients estimated by STE. 
As Figure~\ref{fig:quantization}(b) shows, in LPT, the weights are stored in the format of integers at training, thereby compressing the training memory. 
The model takes the de-quantized weights as input and will quantize the weights back into integers after the backward propagation. 
Since the input one-hot vectors of DLRMs are highly sparse, only extremely small part of the embeddings will be de-quantized into floating-point values, whose memory is negligible. 
Xu et al.~\cite{Int16} uses 16-bit LPT on the embedding table without sacrificing accuracy. 
To enhance the compression capability of LPT, Yang et al.~\cite{Mixed-Precision} proposes a mixed-precision scheme where most embeddings are stored in the format of integers, 
only the most recently or frequently used embeddings are stored in a full-precision cache. 
With a small cache, they  achieve lossless compression with 8-bit or even 4-bit quantization. 
Li et al.~\cite{alpt} proposes an adaptive low-precision training scheme to learn the quantization step size for better model accuracy.


\subsection{Discussion}
Low-precision is a simple yet effective way for embedding compression. 
At the inference stage, binarization can reduce the memory usage by $32\times$ and accelerate the inference through Hamming distance. 
However, the binary embeddings usually cause severe accuracy degradation and need to be trained with the guidance of full-precision embeddings, 
which requires more memory usage and computing resources at training.  
In contrast, quantization has a limited compression capability but can achieve a comparable accuracy as the full-precision embeddings. 
Besides, recent quantization approaches for embedding tables can also compress the memory usage at the training stage 
and improve the training efficiency by reducing the communication traffic.

\section{Mixed-Dimension}\label{sec:md}
Embedding tables usually assign a uniform dimension to all the embeddings in a heuristic way, 
and it turns out to be suboptimal in both prediction accuracy and memory usage \cite{NIS}.
As confirmed in \cite{AutoEmb}, a low dimensional embedding is good at handling less frequent features where a high dimensional embedding cannot be well trained. 
Therefore, to boost the model performance, it is important to assign a appropriate dimension to each feature and use mixed-dimension embeddings.

\begin{figure}[h]
\centering
\includegraphics[scale=0.56]{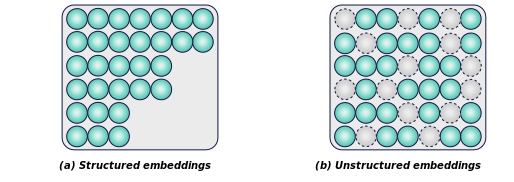}
\caption{Structures of mixed-dimension embeddings.}\label{fig:md}
\end{figure}
Existing methods can obtain mixed-dimension embeddings in a {structured} or an {unstructured} manner. As shown in Figure~\ref{fig:md}, structured approaches divide the embeddings into groups each of which has a unique dimension, while unstructured approaches learn a sparse embedding table where the embeddings have various dimensions. However these mixed-dimension embeddings are not friendly for the operations (e.g., inner product) which require the embeddings of the same length. Therefore, the mixed-dimension embeddings need to be transformed into a uniform dimension before feeding into the following networks. 
Such transformation is usually achieved by linear projection or simply zero padding. Apart from the difference of the embedding structures, existing methods also differ greatly in the way of generating mixed-dimension embeddings. 
In this section, we will introduce three kinds of mixed-dimension approaches named \textbf{rule-based} approaches, \textbf{NAS-based} approaches and \textbf{pruning}, respectively.

\subsection{Rule-based Approaches}
It is a common understanding that the features with higher frequencies are more informative and the fields with more features occupy more memory. 
Thus, the embedding dimension can be set with a heuristic rule based on the feature frequency and the field size.  
To deploy the item embeddings into resource-constraint devices, CpRec~\cite{CpRec} divides the items into several groups by frequency 
and assigns a predefined dimension to each group according to the frequencies of owned features. 
Similarly, MDE~\cite{MDE} assigns each feature field with a unique dimension according to the number of features included in this field. 
Specifically, let $\bm{n} \in \R^m$ denotes the number of features in all $m$ feature fields and $\bm{p} = {1}/{\bm{n}}$, 
then the embedding dimension of $i$-th field would be $ {\bar{d} {\bm{p_i}}^{\alpha}} /{||{\bm{p}}||_{\infty}^{\alpha}}$, 
where $\bar{d}$ is the base dimension and $\alpha \in [0,1]$ denotes the temperature. 
These rule-based approaches are simple yet effective in reducing the memory usage and alleviating the overfitting problems, 
however, they suffer from suboptimal performance as the heuristic rules can not be optimized by the ultimate goal of minimizing the training objective.

\subsection{NAS-based Approaches} \label{sec:nas}
NAS was originally proposed to search for the optimal neural network architectures~\cite{NAS}. 
Recently, it has also been adopted in searching embedding dimensions for different features.  
Unlike the rule-based approaches where the dimension is set based on a priori, it is now learned. 
Generally, there are three components in the NAS-based approaches: 
(1) search space: relaxing the large optimization space of embedding dimensions with heuristic assumptions; 
(2) controller: usually a neural network or learnable parameters, selecting candidate dimension from the search space in a hard or soft manner;
(3) updating algorithm: updating the controller with reinforcement learning (RL) algorithms or differential architecture search (DARTS)~\cite{DARTS} techniques and so on.

We first introduce the approaches of NIS \cite{NIS}, ESAPN \cite{ESAPN} and AutoIAS~\cite{AutoIAS}, which adopt a policy network as the controller and update the controller with RL algorithms.
NIS \cite{NIS} and ESAPN \cite{ESAPN} are designed to learn embedding dimensions of users and items.
In NIS, the authors relax the original search space $n\times d$ to $G\times B$ ($G<n$ and $B< d$) 
by dividing the features into $G$ groups and cutting the embedding dimension of each group into $B$ blocks. 
Then, they use the controller to select several blocks and generate the final embeddings. 
In ESAPN, the authors predefine the search space as a set of candidate embedding dimensions $\mathbb{D}=\{ d_1,d_2,...,d_k\}, $ where $d_1< d_2< ...< d_k$. 
Inspired by the fact that the frequencies of features are monotonically increasing in the data stream, 
they decide to increase or keep the current embedding dimension instead of selecting one in $\mathbb{D}$. 
The decision is made by the controller based on the feature frequency and the current embedding dimension. 
Different from NIS and ESAPN, AutoIAS searches not only the embedding dimensions of feature fields but also the following neural network architectures. 
The authors design a search space for each model component (e.g., the search space of embedding dimensions is similar as $\mathbb{D}$ in ESAPN.). 
To boost the training efficiency, they maintain a supernet at training and use the controller to generate sub-architectures by inheriting parameters from the supernet.

The RL-based approaches described above perform a hard selection by selecting only one embedding dimension for each feature or field at a time. 
Instead, inspired by DARTS, AutoEmb~\cite{AutoEmb} and AutoDim~\cite{AutoDim} make a soft selection by weighted summing over the embeddings of the candidate dimensions in $\mathbb{D}=\{ d_1,d_2,...,d_k\}$. 
Let $\bm{\omega}\in{[0,1]}^k$ denote the vector composed of the weighting coefficients. 
AutoEmb searches for the embedding dimensions of individual features, while AutoDim searches for the embedding dimensions of the feature fields.  
In AutoEmb, the controller is a neural network and generates $\bm{\omega}$ based on the feature frequency. 
While AutoDim directly assigns each field with a learnable vector $\bm{\omega} $, and it further approximates the hard selection by performing gumbel-softmax \cite{Gumbel} on $\bm{\omega}$. 
At training, the controller in AutoEmb and the learnable vectors in AutoDim are optimized through DARTS techniques.  
After training, the corresponding dimension of the largest weight in $\bm{\omega}$ is selected and the model will be retrained for a better accuracy.

Considering that the training process of the controller is quite time-consuming, 
recent works~\cite{Rule, OptEmbed} search for the optimal embedding dimensions after training the models, without using any controller. 
They first sample some structures from the search space and then explore the entire search space by using evolutionary search strategies on the sampled structures. 
Specifically, RULE~\cite{Rule} cuts the embedding table into $G\times B$ blocks similar as NIS and adds a diversity regularization to the blocks in the same group for maximizing expressiveness. 
After training, RULE selects the most suitable embedding blocks under a memory budget (i.e., the maximum number of blocks). 
OptEmbed~\cite{OptEmbed} trains a supernet while removing non-informative embeddings. 
After training, it then assigns each field with a binary mask $\bm{m}\in\{0,1\}^d$ to obtain mixed-dimension embeddings, where $d$ is the original dimension. 
The block selections in RULE and the masks in OptEmbed are determined and evolved by the search strategies. 

\subsection{Pruning}
Instead of shortening the length of embeddings, pruning can obtain a sparse embedding table and thus get mixed-dimension embeddings. 
For instance, DeepLight~\cite{DeepLight} prunes the embedding table in a certain proportion. 
During training, it prunes and retrains the embedding table alternatively so that the mistakenly pruned weights can grow back. 
In addition, DeepLight will increase the pruning proportion gradually as training proceeds. 

Another way to prune the embeddings is to train the embeddings with learnable masks. 
Specifically, an embedding $\bm{e}$ is pruned as $\hat{\bm{e}} =\bm{m}\odot \bm{e}$ for the forward pass, where $\bm{m}$ is the mask and $\odot$ is the element-wise product.
DNIS~\cite{DNIS} divides features into groups by frequency and assigns each group with a learnable mask $\bm{m}\in [0,1]^d $.
AMTL~\cite{AMTL} develops a network to generate a binary mask $\bm{m} \in\{0,1\}^d$ 
for each feature based on its frequency. 
Similarly, PEP~\cite{PEP} generates a binary mask $\bm{m} \in\{0,1\}^d$ for each feature as $\bm{m} = \mathbb{I}(|\bm{e}|> g(s))$, where $\mathbb{I}(\cdot)$ is the indicator function and $g(s)$ (e.g., $sigmoid(s)$) serves as a learnable threshold. 
Specially, in PEP, the embeddings should minus $g(s)$ before being pruned by the masks (i.e. $\hat{\bm{e}}=\bm{m}\odot (\bm{e}-g(s))$). 
At training, the network in AMTL and the learnable threshold in PEP are optimized together with the model parameters by gradient descent, while the learnable masks in DNIS are optimized by DARTS. 
After training, AMTL and PEP preserve $\hat{\bm{e}}$ as the final embeddings, 
while DNIS need pruning $\hat{\bm{e}}$ with a threshold as the redundant weights in $\hat{\bm{e}}$ are not exact zero. The differences between the above methods in generating masks are summarized in Algorithm~\ref{alg:pruning}.

\begin{algorithm}[htbp]
  \SetAlgoLined
  \DontPrintSemicolon
  \KwIn{the full-precision embedding $\bm{e}$ and feature frequency $f$.} 
  \KwOut{the pruned embedding $\hat{\bm{e}}$.\tcp*[l]{$\hat{e}$ will be fed into following networks.}}
  \SetKwFunction{FA}{DNIS}
  \SetKwFunction{FB}{AMTL}
  \SetKwFunction{FC}{PEP}
  \SetKwProg{Fn}{Func}{:}{}
  
  \Fn{\FA{$\bm{e}$}}{
    $\bm{m} :=[0,1]^d$ \tcp*[r]{$\bm{m}$ is a learnable mask.}
    
    $\hat{\bm{e}}=\bm{m}\odot \bm{e}$ \tcp*[r]{$\bm{m}$ is shared by features with similar frequency.}
  }

  \Fn{\FB{$\bm{e}$}}{
    $\bm{m} :=amtl(f)$ \tcp*[r]{$\bm{m}$ is generated by a network $amtl()$ with the frequency $f$.}
    
    $\hat{\bm{e}}=\bm{m}\odot \bm{e}$
  }

  \Fn{\FC{$\bm{e}$}}{
    $\bm{m} = \mathbb{I}(|\bm{e}|> sigmoid(s))$ \tcp*[r]{$sigmoid(s)$ serves as a learnable threshold.}
    
    $\hat{\bm{e}}=\bm{m}\odot (\bm{e}-sigmoid(s))$
  }
  \caption{Comparison between DNIS~\cite{DNIS}, AMTL~\cite{AMTL} and PEP~\cite{PEP}.~\label{alg:pruning}}
\end{algorithm}

To get rid of the extra training process of optimizing the masks, SSEDS~\cite{SSEDS} develop a single-shot pruning algorithm to prune on the pretrained models. For a pretrained model, SSEDS will prune the columns of the embedding matrix for each field and produce structured embeddings. After training, SSEDS assigns each feature field with a mask and form a mask matrix $\mathbf{M} = \{1\}^{\hat{n} \times d}$, where $\hat{n}$ is the number of fields and $d$ is the original embedding dimension.
Instead of learning $\mathbf{M}$ in the training process, SSEDS use $g_{ij} = {\partial f(\mathbf{M},\mathbf{E})}/{\partial \mathbf{M}_{ij}}$ to identify the importance of $j$-th dimension in $i$-th field, 
where $g_{ij}$ is the gradient of the loss function with respect to $\mathbf{M}_{ij}$. 
Specifically, a larger magnitude of $|g_{ij}|$ means that the corresponding dimension has a greater impact on the loss function. 
Note that all $|g_{ij}|$ can be computed efficiently via only one forward-backward pass. 
Given a memory budget, SSEDS calculates a saliency score for each dimension as $s_{ij}={|g_{ij}|}/{\sum_{i=0}^{\hat{n}}\sum_{j=0}^{d}|g_{ij}|}$ and prunes the dimensions with the lowest saliency scores.

\subsection{Discussion}
Mixed-dimension approaches can alleviate the overfitting problems and obtain better accuracy, 
but usually have worse efficiency at the training and the inference stage. 
At inference, the structured approaches usually suffer from extra computing cost due to the linear transformation 
and the unstructured approaches store the sparse embedding table using \textit{sparse matrix storage}, which will cost extra effort to access. 
At training, NAS-based approaches require extremely long time for searching and pruning usually needs to retrain the pruned models for better accuracy which doubles the training time. 
In contrast, rule-based approaches have little influence on the efficiency and can save memory also at the training stage. However, they cannot achieve the optimal accuracy.

\section{Weight-Sharing}\label{sec:ws}
Low-precision approaches reduce the number of bits in a weight and mixed-dimension approaches reduce the number of weights in an embedding. 
Unlike them, weight-sharing approaches share weights among the embedding table, thereby reducing the actual number of parameters within it. 

Existing weight-sharing approaches usually generate embeddings with several shared vectors. 
In this section, we relax the definition of weight-sharing and formulate a weight-sharing paradigm based on existing approaches. 
Specifically, the embedding generation is formulated as $\bm{e}=\bigcup /\sum_{i = 1}^{s}  \bm{I}^i \times \mathbf{T}^i$, 
where $\bigcup /\sum$ denotes concatenation or summation, $\mathbf{T}^i$ is a matrix composed of shared vectors and $\bm{I}^i$ is an index vector for selecting shared vectors in $\mathbf{T}^i$. 
For ease of expression, we refer to the shared vectors as meta-embeddings and the matrices of shared vectors as meta-tables. 
According to the principle of constructing the index vectors, we introduce three kinds of weight-sharing methods named \textbf{hashing}, \textbf{vector quantization}, and \textbf{decomposition}, respectively. 
  
\subsection{Hashing}
Hashing methods generate the index vectors by processing the original feature id with hash functions. 
For instance, the naive hashing method~\cite{old_hash} compresses the embedding table with a simple hash function (e.g., the reminder function). 
Specifically, given the original size of the embedding table $n\times d$, each feature has an embedding $\bm{e} = \bm{I}\times \mathbf{T}$, 
where $\mathbf{T} \in \R^{m\times d} (m< n)$ and $\bm{I}=\onehot(\id \% m)\in \{0,1\}^m$. 
Note that $\bm{I}\times \mathbf{T}$ is actually achieved by table look-up when $\bm{I}$ is a one-hot vector. 
However, \cite{old_hash} naively maps multiple features to the same embedding. The collisions between features will result in loss of information and drop of accuracy.

\begin{algorithm}[htbp]
  \SetAlgoLined
  \DontPrintSemicolon
  \KwIn{the feature id $(\id \leq n)$.} 
  \KwOut{the generated embedding $\hat{\bm{e}}$.\tcp*[r]{$\hat{e}$ will be fed into following networks.}}
  \SetKwFunction{FA}{QR}
  \SetKwFunction{FB}{MEmCom}
  \SetKwFunction{FC}{BCH}
  \SetKwFunction{FD}{FDH}
  \SetKwProg{Fn}{Func}{:}{}
  
  \Fn{\FA{$\id$}}{
    $\bm{I}^1=\onehot(\id\% m), \ \bm{I}^2=\onehot(\id \mid m)$ \tcp*[r]{${m}$ is a predefined parameter.}
    
    $\hat{\bm{e}}=\bm{I}^1 \times \mathbf{T}^1 + \bm{I}^2 \times \mathbf{T}^2 $ \tcp*[r]{$\mathbf{T}^1 \in \R^{m\times d}$ and  $\mathbf{T}^2 \in \R^{(n \mid m)\times d}$.}
    }

  \Fn{\FB{$\id$}}{
    $\bm{I}^1=\onehot(\id\% m), \ \bm{I}^2=\onehot(\id)$ 
    
    $\hat{\bm{e}}=(\bm{I}^1 \times \mathbf{T}^1) \cdot (\bm{I}^2 \times \mathbf{T}^2) $ \tcp*[r]{$\mathbf{T}^1 \in \R^{m\times d}$ and $\mathbf{T}^2 \in \R^{n\times 1}$.}
    }

  \Fn{\FC{$\id$}}{
    divide the bits within $\id$ into $s$ sub-ids $\{\id_1,...,\id_s\}$.

    $\hat{\bm{e}}=\sum_{i = 1}^{s}  \onehot(\id_i) \times \mathbf{T}$ \tcp*[r]{$\mathbf{T}$ is a shared meta-table.}
  }

  \Fn{\FD{$\id$}}{
    \If{feature $\id$ is frequent} 
    {
  $\hat{\bm{e}}:=\R^{d}$ \tcp*[r]{frequent features have unique embeddings.}
    }
    \Else{
      $\hat{\bm{e}}=\text{QR}(\id)$
    }

  }
  
  \caption{Comparison between QR~\cite{QR}, MEmCom~\cite{MEmCom}, BCH~\cite{Binary} and FDH~\cite{Double}.}~\label{alg:hashing}
\end{algorithm}

To reduce the collisions, existing hashing methods use multiple hash functions to process the feature id.  
They usually maintain multiple meta-tables ($\mathbf{T}^1,...,\mathbf{T}^s$) and generate multiple index vectors as $\bm{I}^i = \onehot(\text{hash}^i(\id))$, 
where $i\in[1,s]$ and $\{\text{hash}^i \}_{i=1}^s$ is a group of hash functions. 
For example, QR~\cite{QR} maintain two meta-tables and use the quotient function and the reminder function to generate two index vectors. 
Similarly, MEmCom~\cite{MEmCom} also maintain two meta-tables ($\mathbf{T}^1 \in \R^{m\times d}$, $\mathbf{T}^2\in \R^{n\times 1}$) 
and generate two index vectors as $\bm{I}^1=\onehot(\id\% m)$, $\bm{I}^2=\onehot(\id)$. 
To better distinguish the features, MEmCom multiplies two meta-embeddings as the final embedding. 
Further, Yan et al.~\cite{Binary} use Binary Code based Hash (BCH) functions to process the feature id at bit level. 
It divides the 64 bits of a feature id into $s$ groups and restructures them into $s$ sub-ids ($\id_1,...,\id_s$). 
Each sub-id corresponds to an index vector (i.e., $\bm{I}^i=\onehot(\id_i)$, $i \in [1, s]$) and obtains a meta-embedding. 
Additionally, to enhance the compression capability, BCH keeps one single meta-table and shares it among all $\mathbf{T}^i, i \in [1, s]$.
Although the hashing methods described above are efficient in memory reduction, they still suffer from accuracy degradation and extra computing cost of the hash functions. 
To alleviate these problems, Zhang et al.~\cite{Double} develop a Frequency-based Double Hashing (FDH) method, which only uses hashing on the features with low frequencies.  
In this way, fewer features need to be processed by the hash function. With a little extra storage for the most frequent features, 
FDH not only improves the prediction accuracy but also the inference efficiency. The difference between the above methods in generating embeddings is reflected in Algorithm~\ref{alg:hashing}.

Instead of generating embeddings with meta-embeddings, LMA~\cite{LMA} and ROBE~\cite{ROBE} use hash functions to map each weight in the embedding table into a shared memory ${M}$. 
For a weight $w_{i,j}$ in the embedding table, they take both $i$ and $j$ as the input of hash functions. 
LMA utilizes locality sensitive hashing (LSH) to map the weights of each embedding to ${M}$ randomly. 
ROBE organizes ${M}$ as a circular array and divides the flattened embedding table (i.e. concatenate all rows) into blocks of size $Z$. 
The head of each block is mapped to ${M}$ randomly and the following weights in the block will be mapped to the position next to the head. 

\subsection{Vector Quantization}
Hashing methods typically get the index vector by processing the feature id with hash functions, which fail to capture the similarity between features themselves~\cite{MGQE}. 
To capture the similarity, vector quantization (VQ) constructs the index vectors through approximated nearest neighbor search (ANNS). 
Specifically, for a feature with an original embedding of $\bm{e}$, VQ gets its index vector as $\bm{I}=\onehot(\argmaxA_k \text{sim}(\bm{e},\mathbf{T}_{k})) \in \{0,1\}^m$, 
where $\mathbf{T}_k$ is the $k$-th meta-embedding in the meta-table $\mathbf{T} \in \R^{m\times d}$ and $\text{sim}(\cdot)$ is a similarity function (e.g., Euclidean distance). 
In other words, VQ takes the original embedding as input and quantizes it into its most similar meta-embedding. 
Note that the meta-table and the meta-embedding are commonly referred to codebook and codeword in recent literature on VQ. Here we use meta-table and meta-embedding for consistency. 

Saec~\cite{Saec} generates a meta-table $\mathbf{T}$ by clustering the most frequent embeddings of a pretrained model and then quantizes each original embedding into a meta-embedding in $\mathbf{T}$. 
However, assigning the same meta-embedding to different features (i.e., collisions) still results in drop of accuracy, even though the features have some similarity.  
In addition, Saec cannot optimize the meta-table together with the original embeddings, which also results in suboptimal accuracy.  
To alleviate the collisions, subsequent works adopt product quantization (PQ)~\cite{PQ} and additive quantization (AQ)~\cite{AQ} to quantize an embedding into multiple meta-embeddings. 
To optimize the meta-table together with the original embeddings, 
researchers usually quantize the original embeddings into meta-embeddings during training and use the meta-embeddings as the input of the following network, where the original embeddings will be optimized through STE~\cite{STE}. 

PQ considers an embedding as a concatenation of several segments (i.e., $\bm{e}=\bigcup_{i = 1}^{s} \bm{e}^i$). 
Each segment $\bm{e}^i$ corresponds to a meta-table $\mathbf{T}^i$. 
At training, an embedding $\bm{e}$ is quantized as $\bigcup _{i = 1}^{s}  \bm{I}^i\times \mathbf{T}^i$, where $\bm{I}^i=\onehot(\argmaxA_k \text{sim}(\bm{e}^i,\mathbf{T}^i_k))$. 
In other words, PQ quantizes each segment into its most similar meta-embedding in the corresponding meta-table. 
After training, the original embeddings are discarded and only the meta-tables are preserved. 
Since the selection of a meta-embedding in a meta-table can be compactly encoded by $\log N$ bits, where $N$ is the size of the meta-table, 
an embedding can now be stored by $s\log N$ bits with the help of the meta-tables. 
Further, MGQE~\cite{MGQE} takes the feature frequency into consideration when using PQ. 
Specifically, it divides the embeddings of items into $m$ groups in ascending order of frequency as $\mathbb{G}=\{\mathbf{E}_1,\mathbf{E}_2,...,\mathbf{E}_m\}$ and defines $\mathbb{N}=\{n_1,n_2,...,n_m\}$ 
where $n_1< n_2< ...< n_m$. The embeddings in $i$-th group can only be quantized into the first $n_i$ meta-embeddings in each meta-table. 
Similarly, xLightFM~\cite{xLightFM} performs PQ in each feature field. 
Considering that the feature fields have various size, xLightFM searches for the optimal size (i.e., the number of meta-embeddings) of the meta-tables for each field. 
The search process is achieved by the DARTS algorithm which is similar as the embedding dimension search in Section~\ref{sec:nas}.

Similar to PQ, AQ considers an embedding as a summation of $s$ vectors: $\bm{e}=\sum_{i = 1}^{s} \bm{e}_i$. 
AQ generates its quantized embeddings by $\sum _{i = 1}^{s}  \bm{I}^i\times \mathbf{T}^i$,
$\bm{I}^i=\onehot(\argmaxA_{k} \text{sim} (\bm{e}-\sum_{v = 1}^{i-1}\mathbf{T}_{k_v}^v,\mathbf{T}_k^i))$ 
where $k_v$ is the index of the selected meta-embedding in the $v$-th meta-table. 
Specifically, the first meta-table takes the embedding $\bm{e}$ as input, and outputs its nearest meta-embedding $\mathbf{T}_{k_1}^1$, 
the second meta-table then quantizes the residual part ($\bm{e}-\mathbf{T}_{k_1}^1$) into $\mathbf{T}_{k_2}^2$ and so on. 
The final output embedding $\hat{\bm{e}}=\sum_{i = 1}^{s}\mathbf{T}_{k_i}^i$. 
LightRec~\cite{LightRec} adopts AQ to compress the item embeddings and uses a pretrained model as a teacher to train the meta-tables effectively. 
LISA~\cite{Linear} utilizes AQ to compress the DLRMs where self-attention is performed for sequence processing.
Note that there is a mass of inner product between embeddings in self-attention which suffer from extremely expensive computing costs.
To alleviates this problem, LISA pre-calculates the inner product between meta-embeddings in the same meta-table and stores the results in a small table after training. 
Then, the inner product of embeddings in self-attention can be calculated by summing the inner product of meta-embeddings which can 
accelerate the inference significantly.

\subsection{Decomposition} 
Hashing and vector quantization use one-hot index vectors to perform a hard selection (i.e., selecting only one meta-embedding) in the meta-table 
and alleviate the collisions between features by maintaining multiply meta-tables.
On the contrary, decomposition approaches make a soft selection by summing over all the meta-embeddings in a  meta-table $\mathbf{T}\in \R^{m\times d}$ with a real-valued index vector $\bm{I}\in \R^m$. 
Due to the wide representation space of the real-valued index vectors, one meta-table is sufficient to resolve the collisions between features. 
Each feature will have a unique index vector stored in the index matrix $\mathbf{{I}}_M\in \R^{n\times m}$ when formulating the decomposition as $\mathbf{E} = {\mathbf{I}}_M \times \mathbf{T}$.  

MLET~\cite{Multi} factorizes the embedding table $\mathbf{E}\in \R^{n\times d}$ in terms of $\mathbf{I}_M\in \R^{n\times m}$ and $\mathbf{T}\in R^{m\times d}$. 
Different from the low-rank decomposition where $m< d$, 
MLET decomposes the embedding table into larger matrices (i.e., $m> d$) at training to ensure a larger optimization space. 
After training, MLET generates the embedding table as $\mathbf{E} = {\mathbf{I}}_M \times \mathbf{T}$ for memory reduction and fast retrieval. 
ANT~\cite{Ant} adopt a better initialization for $\mathbf{T}$ and imposes a sparse constraint on $\mathbf{I}_M$. 
Specifically, ANT initializes the meta-table $\mathbf{T}$ by clustering the embeddings of a pretrained model. 
In addition, to reduce redundancy, ANT use an $\ell_1$ penalty on $\mathbf{I}_M$ and constrain its domain to be non-negative. 
Instead of learning and storing the index vectors at training, 
DHE~\cite{DHE} develop a hash encoder $\mathcal{H}: \mathbb{N} \rightarrow \R^m$ to map each feature id into an index vector $\bm{I}\in \R^{m}$ on the fly. 
Specifically, $\mathcal{H}(x)=[h^1(x),h^2(x),...,h^m(x)]$, where $\{h^i\}_{i=1}^m$ is a group of hash functions. With the hash encoder, DHE can eliminate the storage and optimization of $\mathbf{I}_M$. 
Moreover, considering that the index vectors are deterministic and cannot be optimized, DHE further decomposes the meta-table $T$ into a multi-layer neural network to enhance its expressive ability. 

\begin{figure}[h]
\centering 
\includegraphics[scale=0.18]{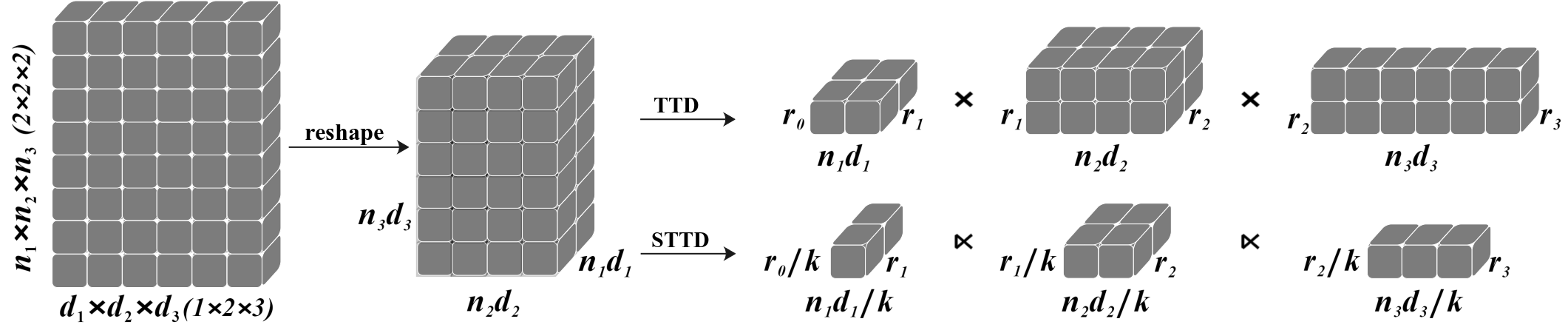}
\caption{Example of TTD and STTD, where $r_3=1$, $r_1=r_2=k=2$, $r_0$ is $1$ in TTD and is $2$ in STTD.}\label{fig:ttd}
\end{figure}

Different from the above naive decomposition, \cite{TT,TT-KD,STT} use tensor train decomposition (TTD) to decompose the embedding tables. 
As shown in Figure~\ref{fig:ttd}, the embedding table $\mathbf{E}\in \R^{n\times d}$ will first be reshaped into 
$\mathcal{E}\in \R^{(n_1 d_1)\times (n_2 d_2)\times ...\times (n_s d_s)}$, where $n=\prod_{i=1} ^s n_i$ and $d=\prod_{i=1} ^s d_i$. 
Then, $\mathcal{E}$ will be decomposed as $\mathcal{E}=\mathcal{G}_1\times\mathcal{G}_2\times...\times\mathcal{G}_s$, 
where $\mathcal{G}_i\in\R^{r_{i-1}\times n_id_i\times r_{i}}$. $\{\mathcal{G}_i\}_{i=1}^s$ are called TT-cores and $\{r_i\}_{i=0}^s$ are called TT-ranks, in particular $r_0=r_s=1$. 
TT-Rec~\cite{TT} is the first to use TTD on the embedding tables of DLRMs. It implements optimized kernels of TTD for embedding tables. 
LLRec~\cite{TT-KD} uses TTD on the embedding tables while maintaining the prediction accuracy by knowledge distillation. 
To enhance the compression capability of TTD, \cite{STT} further develop semi-tensor product based tensor train decomposition (STTD). 
Semi-tensor product is a generalization of matrix product. Specifically, given $\bm{a}\in\R^{1\times np}$ and $\bm{b}\in\R^{p}$, $\bm{a}$ can be cut into $p$ blocks $\{\bm{a}^i \in \R^{1\times n} \}_{i=1}^p$ 
and $\bm{a} \ltimes \bm{b} = \sum_{i=1}^p \bm{a}^i\times \bm{b}_i \in \R^{1\times n}$, where $\ltimes$ is the left semi-tensor product.
For matrices $\mathbf{A}\in\R^{h\times{np}}$ and $\mathbf{B}\in\R^{p\times q }$, $\mathbf{A}\ltimes\mathbf{B} \in \R^{h\times nq}$ contains $h\times q$ blocks 
and each block is the semi-tensor product between a row of $\mathbf{A}$ and a column of $\mathbf{B}$. \cite{STT} replaces the conventional matrix tensor product of TTD with the left semi-tensor product. 
As Figure~\ref{fig:ttd} shows, in STTD, $\mathcal{E}=\mathcal{\hat{G}}_1\ltimes\mathcal{\hat{G}}_2\ltimes...\ltimes\mathcal{\hat{G}}_s$, 
where $\mathcal{\hat{G}}_i \in \R^{\frac{r_{i-1}}{k}\times \frac{n_id_i}{k} \times r_i}$ ,$r_0=k$ and $r_s=1$. 
In addition, \cite{STT} uses self-supervised knowledge distillation to to reduce accuracy loss from compression.

\subsection{Discussion}
Weight sharing approaches usually make remarkable reduction to the memory usage. 
However, they suffer from low efficiency at training due to extra computing cost for generating embeddings, 
especially the nearest neighbor search in vector quantization and the matrix multiplication in decomposition approaches.  
The extra computing cost will also slow down the inference speed except in vector quantization where we can store the results of inner product between meta-embeddings to accelerate the inference. 
Nevertheless, vector quantization maintains the original embeddings during training which requires extra memory usage. 
Moreover, these methods usually cannot improve the prediction accuracy, especially hashing usually causes severe drop of accuracy.

\section{Summary} \label{sec:summary}
Embedding tables usually constitute a large portion of model parameters in DLRMs, which need to be compressed for efficient and economical deployment. 
As recommender systems continue to grow in scale, embedding compression has attracted more and more attention. 
In this survey, we provide a comprehensive review of the embedding compression methods in recommender systems, accompanied by a systematic and rational organization of existing studies. 

The embedding table can be conceptualized as a matrix with three dimensions, namely the precision of weights, the dimension of embeddings, and the number of embeddings. 
Consequently, we classify embedding compression methods into three primary categories according to the dimensions they compress, which are low-precision, mixed-dimension, and weight-sharing, respectively. 
Low-precision methods reduce the memory of each weight by decreasing its bit width, including binarization and quantization. 
Mixed-dimension methods reduce the memory of specific embeddings by decreasing their dimensions, including rule-based approaches, NAS-based approaches and pruning. 
Weight-sharing methods reduce the actual parameters of the embedding table by sharing weights among different embeddings, including hashing, vector quantization and decomposition. 

\subsection{General Suggestions}
In the above sections, we have discussed the pros and cons of different compression methods in detail. However, there are no golden criteria to measure which one is the best.  
How to choose a proper compression method depends greatly on the application scenarios and requirements. 
Therefore, we offer some general suggestions for the common requirements on the key metrics discussed in Section~\ref{sec:emb_comp}, namely model accuracy, inference efficiency, training efficiency, and training memory usage, respectively. 

\begin{itemize}[leftmargin=*]
\item \textbf{Model accuracy.} 
In scenarios that demand high model accuracy, any accuracy degradation caused by compression is deemed unacceptable. 
In such cases, mixed-dimension methods are recommended as they have been reported to remove redundant parameters and avoid model overfitting. With an appropriate compression ratio, mixed-dimension methods can effectively compress embeddings while maintaining or even improving accuracy. 
Furthermore, accuracy can also be preserved when compressing embeddings by quantization with a higher bit width. For instance, using a 16-bit representation has been proven to be sufficient for achieving accurate results.
On the contrary, for scenarios that do not require high prediction accuracy, quantization with lower bit width or even binarization (1-bit) can be employed to achieve stronger compression. 
\item \textbf{Inference efficiency.} 
In scenarios such as online inference, model inference efficiency is of paramount importance. Generally speaking, most embedding compression methods will not have a great negative impact on the inference speed. However, in several decomposition methods where the embedding table is decomposed into multiple small matrices, the process of recovering embeddings may introduce significant inference latency and should be avoided in this context. 
To improve inference efficiency while compressing embeddings, vector quantization is suggested, as the feature interaction (e.g., inner-product) of embeddings can be pre-calculated to accelerate the inference process. Additionally, binarization is also worth considering when there is no high requirement on model accuracy. The calculation of feature interactions between binary embeddings is faster compared to that between full-precision embeddings. 
\item \textbf{Training efficiency.} 
In scenarios where the models are supposed to be updated in a timely manner, training efficiency becomes a critical factor. However, it is unfortunate that most embedding compression methods do not contribute to improving the training efficiency. In fact, some of them may significantly reduce training efficiency, particularly NAS-based approaches, pruning, vector quantization, and decomposition.
Specifically, NAS-based approaches involve complex calculations to search for optimal embedding dimensions, which can be computationally intensive and time-consuming. Pruning often necessitates retraining to achieve higher accuracy, resulting in additional training overhead. Vector quantization also involves cumbersome calculations for nearest neighbor searches. Decomposition may require multiple matrix multiplications to recover and retrieve embeddings. 
Therefore, in scenarios that prioritize training efficiency and timely model updates, these methods are not recommended.
\item \textbf{Training memory usage.} 
In scenarios where computing devices have limited memory, it is desirable to compress the training memory usage of embeddings or, at the very least, avoid increasing it.
In such cases, we suggest using rule-based approaches, hashing, or decomposition, as they can compress the embedding table before training. Besides, the low-precision training of quantization is also worth considering, as the embeddings are stored in the format of integers during training. 
On the contrary, NAS-based approaches and vector quantization are not recommended in this context. They often require storing a significant number of intermediate results to guide the training process, which will consume more memory.

\end{itemize}

\subsection{Future Prospects}
Embedding compression in recommender systems has witnessed rapid development and notable achievements, although there are still several challenging issues that require attention. We identify several potential directions for further research in this field.

\begin{itemize}[leftmargin=*]
\item \textbf{Low-precision.}
The key problem faced by low-precision methods is the severe accuracy degradation at extremely lower bit widths. 
In view of the extensive and advanced research on quantization and binarization in the deep learning community, we can refer to related techniques to alleviate the accuracy loss when compressing embeddings, which is quite challenging and valuable. 
\item \textbf{Mixed-dimension.}
In recent advanced mixed-dimension methods, there is a need to enhance the training efficiency of NAS-based approaches and pruning. 
To address this, we recommend designing lighter NAS frameworks that can efficiently search for the optimal embedding dimension. 
On the other hand, finding solutions to avoid retraining pruned models is also crucial for enhancing training efficiency. 
Furthermore, while numerous studies have demonstrated the significant impact of the embedding dimension on model accuracy, there is still a lack of theoretical understanding regarding how the embedding dimension precisely affects model accuracy. Having a solid theoretical basis would be invaluable in guiding the optimal selection of embedding dimensions, enabling more efficient and effective model training.
\item \textbf{Weight-sharing.} 
To approach the limit of weight sharing methods, we believe that an intriguing direction to explore is the use of embedding generation networks. 
Considering the powerful representation capabilities of neural networks, we may learn a powerful neural network to generate embeddings, instead of directly  learning and maintaining the embeddings themselves. 
\item \textbf{Hybrid approaches.} 
Since the methods within the three primary categories compress the embeddings from different dimensions and enjoy different advantages, we expect future research to establish a unified method for compressing multiple dimensions, or develop hybrid approaches combining these techniques. By integrating the strengths of different compression methods, it is possible to create more powerful and comprehensive compression algorithms. 
\item \textbf{Open benchmarks.} 
This review offers a thorough discussion of embedding compression methods. However, we did not undertake an experimental comparison across these methods. On one hand, distinct methods are applied to different tasks in recommender systems, each of which has unique accuracy metrics. For example, in click-through rate (CTR) prediction, the commonly used metric is the Area Under the Curve (AUC); whereas for rating prediction, Root Mean Square Error (RMSE) and Mean Absolute Error (MAE) are typically employed; for Top-N recommendations, Mean Average Precision (MAP) and Normalized Discounted Cumulative Gain (NDCG) are commonly utilized as accuracy metrics. On the other hand, a majority of research relies on proprietary datasets without sharing open-source code, presenting obstacles to reproducibility and comparative analyses. Nonetheless, the implementation of these methods is not inherently complex. Given the focus on the embedding tables, a solution involves the definition of a new embedding module during implementation, coupled with the rewriting of the lookup operation for the embedding vector. Therefore, it is necessary to establish a foundational benchmark to evaluate the effectiveness of distinct methods across a spectrum of tasks, like BARS~\cite{bars}, a benchmark designed for recommendations. We posit that this would substantially expedite the application and advancement of this field.
\end{itemize}

\section*{Acknowledgments}
This work is supported in part by National Natural Science Foundation of China under grants 62376103, 62302184, 62206102, and Science and Technology Support Program of Hubei Province under grant 2022BAA046.

\bibliographystyle{ACM-Reference-Format}
\bibliography{reference}
\end{document}